\documentclass{iau}
\usepackage{graphicx}

\newcommand{\atlas}{{{ATLAS$^{\mathrm{3D}}$}}}
\newcommand{\sauron}{{\tt {SAURON}}}

\newcommand{\lsim}{\mathrel{\hbox{\rlap{\hbox{\lower4pt\hbox{$\sim$}}}\hbox{$<$}}}}
\newcommand{\gsim}{\mathrel{\hbox{\rlap{\hbox{\lower4pt\hbox{$\sim$}}}\hbox{$>$}}}}
\newcommand{\msun}{$M_{\odot}$}
\newcommand{\nmsun}{$\mathrm{M}_{\odot}$}

\title[IAU Symposium 295.~~The stellar populations of massive galaxies in the local Universe] 
{The stellar populations of massive galaxies \\ in the local Universe}

\author[Richard M. McDermid]   
{Richard M. McDermid$^1$
}

\affiliation{$^1$Gemini Observatory Northern Operations Centre, \\ 670 N. A'ohoku Pl.,
Hilo HI 96720, USA \\ email: {\tt rmcdermid@gemini.edu} 
}

\pubyear{2013}
\volume{295}  
\pagerange{xx--xx}
\jname{The intriguing life of massive galaxies}
\editors{D. Thomas, A. Pasquali \& I. Ferreras, eds.}
\begin{document}

\maketitle

\begin{abstract}
I present a brief review of the stellar population properties of massive galaxies, focusing on early-type galaxies in particular, with emphasis on recent results from the \atlas\/ Survey. I discuss the occurrence of young stellar ages, cold gas, and ongoing star formation in early-type galaxies, the presence of which gives important clues to the evolutionary path of these galaxies. Consideration of empirical star formation histories gives a meaningful picture of galaxy stellar population properties, and allows accurate comparison of mass estimates from populations and dynamics. This has recently provided strong evidence of a non-universal IMF, as supported by other recent evidences. Spatially-resolved studies of stellar populations are also crucial to connect distinct components within galaxies to spatial structures seen in other wavelengths or parameters. Stellar populations in the faint outer envelopes of early-type galaxies are a formidable frontier for observers, but promise to put constraints on the ratio of accreted stellar mass versus that formed `in situ' - a key feature of recent galaxy formation models. Galaxy environment appears to play a key role in controlling the stellar population properties of low mass galaxies. Simulations remind us, however, that current day galaxies are the product of a complex assembly and environment history, which gives rise to the trends we see. This has strong implications for our interpretation of environmental trends.

\end{abstract}

\firstsection 
\section{Introduction}

Much of what has been written about the stellar population content of nearby massive galaxies has focussed on that of early-type galaxies specifically, by obvious virtue of their observational simplicity, having relatively high surface brightnesses, simple morphologies, and very little dust or emission to complicate the interpretation of their spectra. I will not deviate from this well-worn path in this review, but point out that `massive galaxies' need not be synonymous with early-type galaxies alone, which only dominate the stellar mass function above 3.e10~\msun\/ \cite[(e.g. Baldry et al. 2006)]{baldry06}. That said, a large portion of the stars in the local Universe do exist in early-type galaxies, mostly in the form of old stellar populations that carry the fossil evidence of earlier evolutionary epochs. In this review, I focus on the properties of early-type galaxies, with a bias towards recently published results from the \atlas\/ Survey\footnote{http://www.purl.org/atlas3d} \cite[(Cappellari et al., 2011)]{cappellari11}.

\vspace{-0.5cm}
\section{Colour bimodality}

The colour-magnitude diagram of galaxies \cite[(in particular that from SDSS, e.g. Kauffman et al.,  2003, Baldry et al., 2004)]{kauffman03, baldry04} is an excellent tool to summarize our current knowledge and ideas of galaxy evolution, being a snapshot of the cumulative star formation history (color) and mass assembly (magnitude) for the population. In an evolutionary context, it is inevitable to ask how galaxies may move around in this parameter space as a function of time, and what processes give rise to the main features of the distribution.

Within the hierarchical framework of $\Lambda$CDM, we assume that galaxies largely follow their host dark matter haloes, and increase in mass over time. Likewise, the stellar populations will evolve over time, becoming redder. New stars may form if sufficient fuel exists and conditions allow it, producing or preserving bluer colors. Various tracks through the color-magnitude distribution have been proposed, most of which consider that blue galaxies are the building blocks for most red-sequence galaxies, with evolution {\it along} the red sequence also necessary to produce the most massive galaxies, which have insufficient blue counterparts of similar mass. The gradual reddening process of gas exhaustion and stellar evolution, however, do not obviously give rise to the observed bimodal color distribution. Creating an accelerated transition from blue to red colors can be achieved with the addition of a mechanism which `quenches' star formation \cite[(Faber et al., 2007)]{faber07}.

\vspace{-0.5cm}
\section{Feedback}

One suggested mechanism for quenching star formation is that of energy input into the star-formation gas reservoir. The form that this energy takes is not currently clear, nor is the precise source of energy input (AGN or stellar winds). Circumstantial evidence exists, however, to support the cessation of star formation due to the onset of black hole accretion. Specifically, the work of \cite[Schawinski et al. (2007)]{schawinski07} found that the most active AGN galaxies lie in the intermediate region between red and blue galaxies but have only small amounts of recently-formed stars, strongly suggesting that the AGN activity is associated with the quenching of star formation.

An example of a {\it direct} connection between nuclear accretion and the removal of star formation fuel comes from the recent discovery of a massive molecular gas outflow from the early-type galaxy NGC\,1266 \cite[(Alatalo et al., 2011, Davis et al., 2012)]{alatalo11,davis12}. This otherwise typical lenticular galaxy shows molecular gas flowing away from the central regions faster than the escape velocity, losing gas faster through the outflow than via the apparent star formation rate. X-ray and radio signatures suggest the presence of an AGN, which would be an obvious power source. This object appears somewhat unique within the volume-limited parent sample of 260 early-type galaxies it was taken from, observed as part of the \atlas\/ Survey. However, it is a bright and easily detected case, and less obvious examples may be lurking below the sensitivity limits of current surveys.

\vspace{-0.5cm}
\section{Cold gas in early-type galaxies}
\label{sec:gas}

\begin{figure}[t]
\begin{center}
 \includegraphics[width=5.5in]{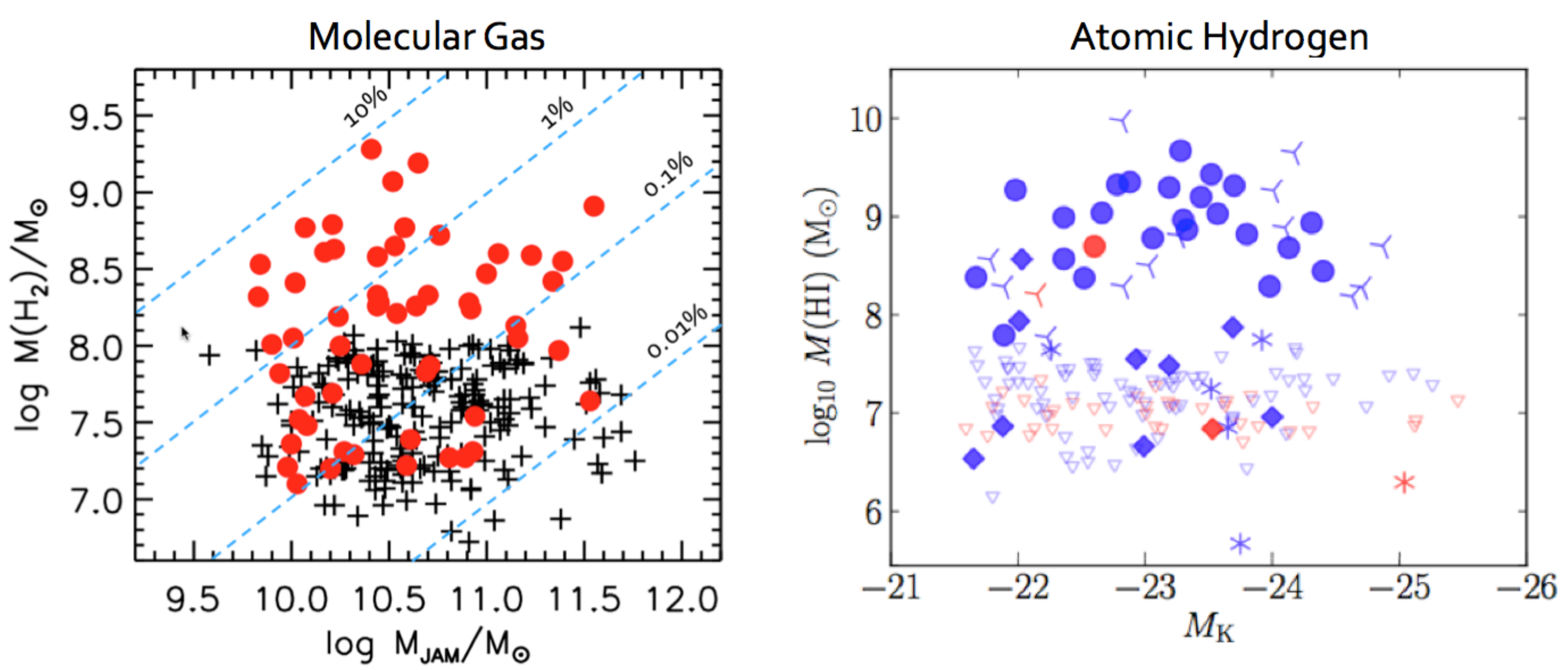} 
 \caption{{\it Left}: Molecular gas masses for the \atlas\/ sample as a function of dynamical mass, from \cite[Young et al. (2011)]{young12}. Circles and crosses are detections and non-detections respectively. {\it Right}: Neutral gas mass for \atlas\/ galaxies as a function of absolute K-band magnitude, from \cite{serra12}. Filled and open symbols are detections and non-detections respectively.}
   \label{fig1}
\end{center}
\end{figure}

The largest and most sensitive surveys targeting early-types in CO (single dish, J=1-0, 1-2) and HI (interferometric, 21~cm) were recently completed as part of the \atlas\/ Survey \cite[(Young et al., 2011, Serra et al., 2012)]{young11,serra12}, building on previous investigations \cite[(Morganti et al., 2006, Young et al., 2008)]{morganti06,young08}. By virtue of being a complete sample, \atlas\/ has provided secure detection fractions for HI and CO down to sensitive mass limits (around $10^7$~\msun). Both tracers show numerous cases where the gas mass and gas fraction are comparable to that of spiral galaxies. More than 20\% of early-type galaxies show detectable cold-gas content in either tracer, with a strong correspondence of HI and CO in the same objects, and little dependence on galaxy mass (see Figure 1).

CO interferometry was obtained for the 30 brightest detections in the \atlas\/ Survey \cite[(Alatalo et al., 2012)]{alatalo12}, allowing the spatial distribution of the CO gas to be investigated. The extent and star-gas kinematic alignment derived from these data have revealed interesting constraints on the origin of the molecular gas in these objects, implying important differences in the molecular gas properties in different environments \cite[(Davis et al., 2011, see also Davis et al. this volume)]{davis12}. Specifically, objects in the Virgo cluster (the densest environment of the survey) almost exclusively show aligned, co-rotating CO gas, unlike the field which shows a similar occurrence of aligned and misaligned gas kinematics. This supports a picture where field galaxies continue to accrete gas via minor mergers with in falling systems, giving a range of resulting orientations, whereas in the cluster environment, this channel is apparently suppressed, and an internal origin seems likely.

Detection rates of CO in Virgo are also as high as in the field \cite[(Young et al., 2011)]{young11}. This contrasts with the HI detection rate, which is almost nil within the cluster, suggesting that the denser CO gas is more robust to the tidal interactions and ram pressure of the cluster environment than the HI. HI morphology also shows a great variety, with disks, resonant rings, and tidally-disturbed structures, all of which can exist on 10s of kpc scales \cite[(Serra et al., 2012)]{serra12}. The CO, by contrast, is typically found in regular rotating disk structures within the central few kpc \cite[(Alatalo et al., 2012, Davis et al., 2012)]{alatalo12,davis12}.

\vspace{-0.5cm}
\section{Recent and ongoing star formation in red-sequence galaxies}

The presence of neutral gas, and in particular molecular gas, at levels of up to 10\% by mass is a strong indication that these objects are currently experiencing star formation. This is consistent with findings from the \sauron\/ Project, which looked for excess flux in the $8 \mu$m Spitzer/IRAC band, indicative of PAH emission from star formation \cite[(Shapiro et al., 2010)]{shapiro10}. The rates of star formation found are typically low (around 0.1~\nmsun/yr) with some objects getting closer to spiral-like values (1~\nmsun/yr). The star formation rate does not depend strongly on galaxy mass, although the {\it morphology} of the star-forming region does. More massive galaxies generally show star formation within disks that have small scale lengths compared to the half-light radius of the stars, whereas going to lower mass galaxies, the star formation region can be as extended as the stellar light. This creates very different impacts on the color and, more sensitively, the mean luminosity-weighted age of the galaxy.

Evidence of young stars in early-type galaxies comes from the first detailed studies of stellar populations using absorption lines and population synthesis models 
 \cite[(e.g. Gonzalez, 1993; Trager et al., 2000)]{gonzalez93,trager00}. Young mean ages inferred from integrated optical light is straightforwardly interpreted as a `frosting' of young stars superimposed on a body of mostly old stars. Later studies refined the picture to show, for large samples, that young galaxy ages were more common in objects with lower velocity dispersion \cite[(Thomas et al., 2005)]{thomas05}, and indeed the two parameters are tightly correlated \cite[(e.g. Graves et al., 2009)]{graves09}.

The presence of young stars is most easily seen at bluer wavelengths, making the ultra violet (UV) a particularly sensitive tracer of stars formed in the past Gyr. \cite{kaviraj07} showed the dramatically increased dispersion of the red sequence when using UV-optical colors, driven primarily by the presence of young ($\lsim 1$\,Gyr) stars in more than 30\% of the present day early-type galaxy population.

\vspace{-0.5cm}
\section{Star formation histories of early-type galaxies}


As high-quality stellar libraries and associated stellar population models have become available \cite[(e.g. S{\'a}nchez-Bl{\'a}zquez et al., 2006; Vazdekis et al. 2012)]{sanchez08,vazdekis12}, it is now possible to fit sizable optical-NIR wavelength ranges at moderate spectral resolution with spectra that each represent a population of fixed age and metallicity, with additional free parameters like the assumed initial mass function and elemental abundances \cite[(Conroy \& van Dokkum, 2012)]{conroy12}. This allows the full information content of the spectrum to be used, and avoids changing the observed data to accommodate the specific model framework by instead adapting the models to the data. 

A number of approaches have been employed to use such models and spectral fitting to derive the star formation histories of galaxies, typically either selecting a fixed functional form (such as an exponential), and solving for the function's free parameters \cite[(e.g. Pforr et al. 2012)]{pforr12}; or selecting the best fitting scenario from a large library of different trial histories, which can take arbitrary forms \cite[(Aquaviva et al., 2011)]{aquaviva11}. Another approach is to take a regularized linear combination of model components and find the `smoothest' distribution of free parameters that can still adequately fit the data \cite[(Ocvirk et al., 2006)]{ocvirk06}.


\vspace{-0.5cm}
\section{Galaxy mass-to-light ratios}


As well as being instructive from a galaxy evolution perspective, knowing the distribution of population parameters within a galaxy can have a large impact in how we interpret the stellar light in terms of mass, via the stellar mass to light ratio. For example, a galaxy may have absorption line strengths that are fully consistent with a single stellar population (SSP) age of 3~Gyr. The same combination of indices may be equally well fitted by combination of two different SSPs, say 95\% by mass of 13\,Gyr and 5\% of 1\,Gyr. These two scenarios would differ in inferred mass by almost a factor of 3.

This effect was shown in \cite{cappellari06} where the population and dynamical M/L properties of 24 early-type galaxies, all with high-quality integral-field spectroscopy from the \sauron\/ Survey, were compared. Younger galaxies deviated systematically from the identity line towards lower population-based M/L. By modeling these younger galaxies using two population components, one older (12~Gyr, 90\% by mass), and one younger, the dynamical- and population-based M/L values came into agreement.

An analogous study was recently made by \cite[Cappellari et al.\,(2012a, 2012b)]{cappellari12a,cappellari12b} using the \atlas\/ Survey of 260 early-type galaxies, again modeling the stellar dynamics and populations derived from integral-field spectroscopy covering to around one effective radius for each galaxy. Star-formation histories were derived for each galaxy using the regularized spectral fitting method of \cite{cappellari04}, fitting for the age and metallicity distribution in each object. From these distributions of mass contributions in each age and metallicity bin, it is possible to compute the mass in stars, as well as the light output, giving effectively the mass-weighted M/L.

For the dynamical M/L estimates, it is first necessary to describe the stellar light distribution. It is important to do this as accurately as possible, as it will determine the assumed stellar mass distribution, which will in turn affect the spatially resolved kinematics predicted for a given M/L. Early-type galaxies show a wealth of detailed structure in their spatially resolved surface brightness distributions, even though they are often assumed to follow computationally convenient, spherically-symmetric profiles. Instead,  the 260 SDSS r-band galaxy images were fitted in detail using the Multi-Gaussian Expansion (MGE) approach \cite[(Monet et al., 1992; Emsellem et al., 1994)]{monet92,emsellem94}. Even imposing the constraint of axisymmetry, the light distribution is fitted to high precision, matching substructures such as disks, boxiness, and variations in radial profile shapes that are missed in simpler models.

The dynamical M/L estimates were computed using the Jeans Anisotropic MGE (JAM) approach of \cite{cappellari08}, which extends the semi-isotropic ($f = f(E, L_z)$, two integral) formalism of \cite{emsellem94} to the case where the velocity ellipsoid is aligned with the cylindrical coordinates $(R, z)$, and the anisotropy is parameterised as $\beta_z = 1-\overline{v_{z}^{2}}/\overline{v_{R}^{2}}$. The free parameters are the M/L, $\beta_z$, and the inclination of the system, $i$. Under these assumptions, the Jeans equations are solved to produce an estimate of the projected second velocity moment $V_{rms}=\sqrt{v^2 + \sigma^2}$ as a function of position, which is compared to the observed kinematics, and the best-fit case found. The resulting fits to the high-quality kinematic data tightly constrain the M/L. 

\vspace{-0.5cm}
\section{Measuring variations in the IMF}

\begin{figure}[t]
\begin{center}
 \includegraphics[width=5.5in]{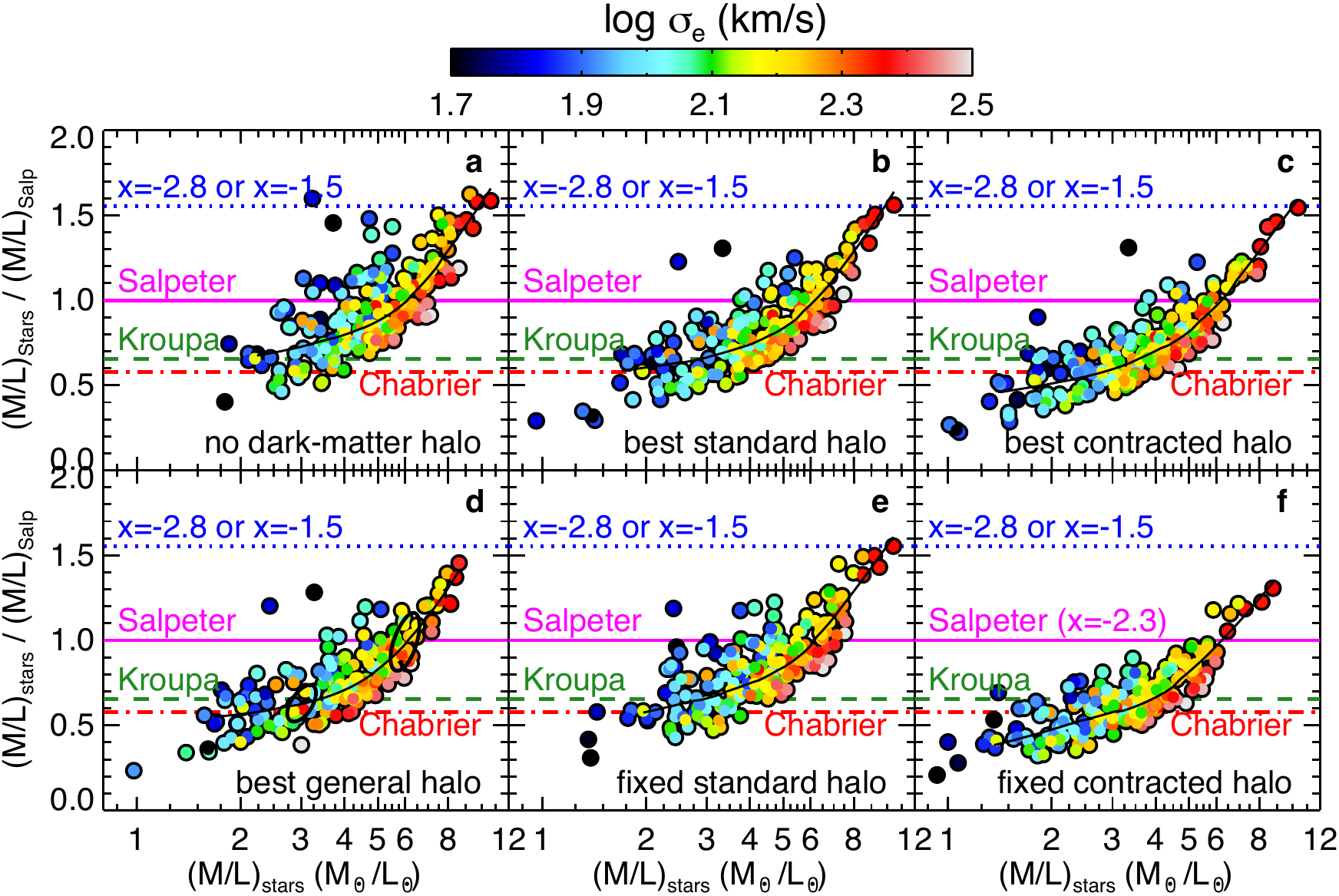} 
 \caption{Ratio of stellar M/L derived from stellar dynamical modeling and stellar population models, plotted as a function of dynamical M/L. Symbol colours correspond to velocity dispersion. Horizontal lines indicate where the points would lie for different canonical IMFs. The six different panels illustrate the best fitting results under different dark matter assumptions, as indicated in each panel. In all cases, the ratio of the two M/L estimates is not constant, and shows the same systematic deviation towards higher mass normalisations in higher mass galaxies. From \cite[Cappellari et al.\,(2012a)]{cappellari12a}.}
   \label{fig2}
\end{center}
\end{figure}

In addition to the effects of young stellar populations on the inferred M/L, \cite{cappellari06} also showed that, adopting a universal form of the stellar initial mass function (IMF), it is not possible to have the M/L from populations and dynamics agree for all galaxies. Specifically, there was a weak trend that higher-mass galaxies showed increasing dynamical M/L with no corresponding increase in population M/L - i.e. the stellar kinematics showed a mass component in these galaxies that is unaccounted for in the light. Two possibilities are: (1) that the dark matter fraction is different in the central parts of these galaxies; or (2) the initial mass function of the stellar populations is different. However, the effects of these two possibilities are degenerate for the data and modeling employed. Interestingly, this same finding was later confirmed using different data and an independent approach, via gravitational lensing \cite[(Treu et al., 2010)]{treu10}.

With \atlas, the sample size increases ten-fold over the \sauron\/ study, with comparable 2D kinematic and imaging data. The JAM modeling approach is computationally much more efficient compared with more general methods
, and is readily applied to large samples. Integral field data are also essential here, as the inclination and anisotropy vary the {\it shape} of the $V_{rms}$ distribution away from the principle axes, which would not be seen with major- or minor-axis long slit data.

The fit to the stellar light and 2D $V_{rms}$ is sufficiently constraining that, for the first time, it is possible to break the degeneracy between the stellar mass distribution and the dark matter component, thus allowing trends in these properties to be tracked across the galaxy sample. This was done by assuming several different models for the distribution of the dark matter, each motivated by current theoretical models. Figure \ref{fig2} shows the results.
If the estimates from both methods were consistent, the points would all lie horizontally. Since the dark matter variations are accounted for, this variation must come from the initial mass function of the stellar population (which is held fixed in this diagram).

This finding supports recent claims (van Dokkum \& Conroy, 2010, Conroy \& van Dokkum 2012) that show spectral evidence for large numbers of faint, low-mass dwarf stars in the centers of early-type galaxies. This claim has been generally supported by numerous other similar studies \cite[(e.g. Spiniello et al., 2012; Ferreras et al., 2012; Smith et al., 2012b)]{spinielo12,ferreras12,smith12b}. We note that these spectral studies are required to distinguish the slope of the IMF, as gravitational studies alone cannot distinguish between dark mass from faint low-mass stars, or remnants from massive star evolution.

\vspace{-0.5cm}
\section{Spatially-resolved stellar populations}


The variations of stellar populations {\it within} massive galaxies indicate the diversity of processes in their evolution, and are consistent with a number of possible formation scenarios. Once again, the power of integral-field spectroscopy gives a unique view of these properties, allowing spatial structures in the populations to be unambiguously associated to other features, such as kinematic components, dust lanes, gas disks, etc. In this way, \cite{kuntschner10} showed that the presence of young stars is often accompanied by the presence of a rapidly-rotating disk component in the kinematics. In some cases, these components also show disks of molecular gas that are co-spatial with the young stellar disk \cite[(Young et al., 2008)]{young08}, showing that gas has settled onto the equatorial plane of the system and is forming new stars.

Metallicity gradients are essentially ubiquitous in early-type galaxies of all masses, and occur in the sense that metallicity decreases in the outskirts, falling approximately as a power law with radius. A more physical interpretation of these gradients comes from considering that the metallicity is tracing a change in the gravitational potential, which can be characterized by the escape velocity. A recent study by \cite{scott09} has shown that early-type galaxies collectively follow trends in there local stellar population -- escape velocity relations. This is suggestive that the formation of metals does not care about the specific galaxy as much as what the local potential is. In a situation where stellar winds control the star-formation process, deeper potentials could retain more enriched gas, leading to a strong connection between the metallicity and the local potential. Dry galaxy merging would disrupt this connection, and so the tightness of the local-global trends limits the number of major dry mergers to approximately 1.5 over a Hubble time \cite[(Scott et al. 2009)]{scott09}.

\begin{figure}[t]
\begin{center}
 \includegraphics[width=5.5in]{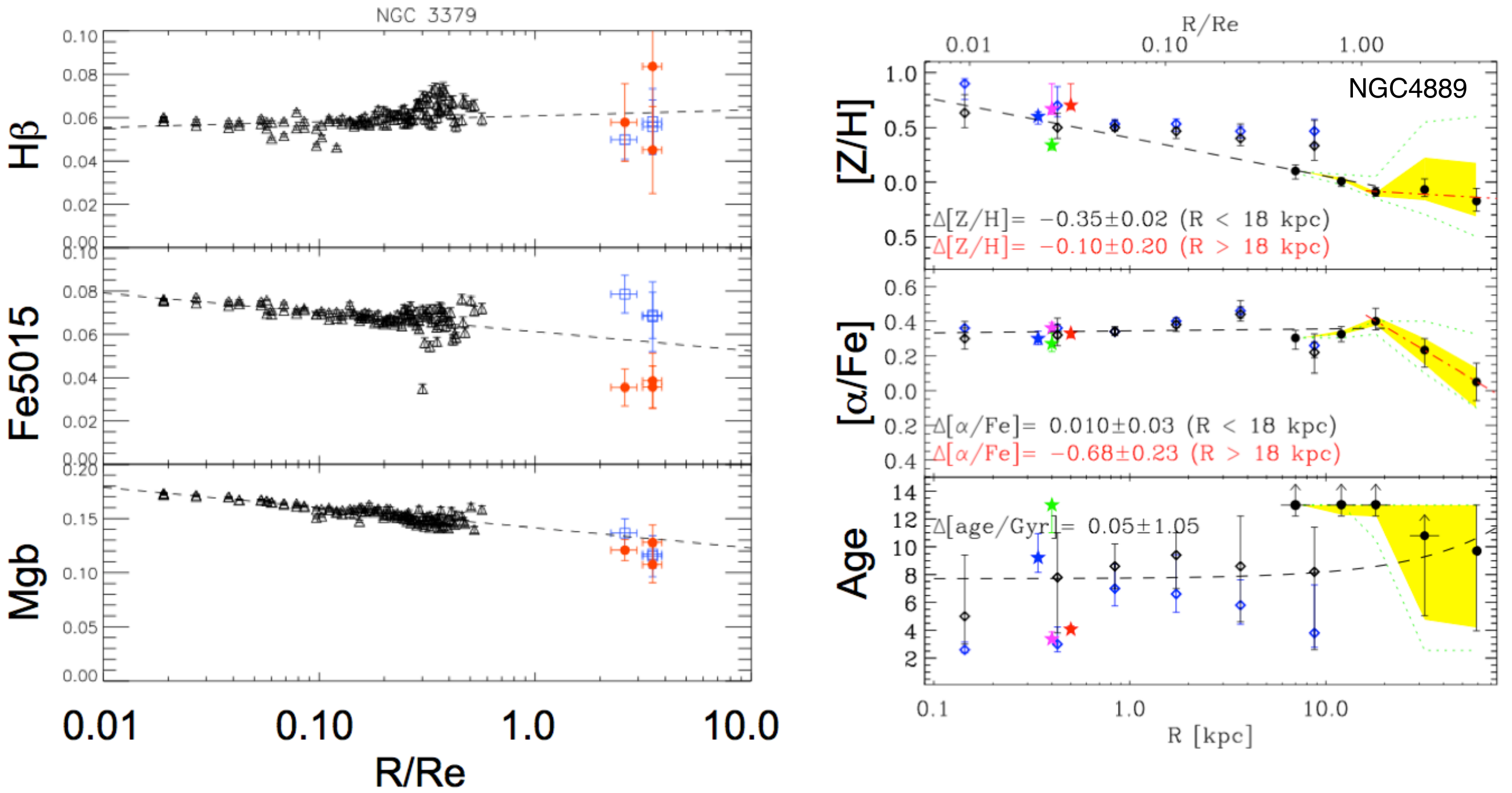} 
 \caption{{\it Left}: Radial line-strength gradients for NGC\,3379, from \cite{weijmans09}. {\it Right}: Radial profile of SSP parameters for NGC\,4889, from \cite{coccato10}.}
   \label{fig3}
\end{center}
\end{figure}

A somewhat different scenario for building metallicity gradients in was proposed by \cite{oser10}, based on simulations. Here the central regions of the most massive galaxies form very early on, through rapid gas inflow and subsequent collapse, quickly forming a moderately massive, compact object. These then proceeded to accrete lower mass galaxies. The merging process is such that most of the stars of an accreted satellite infall only as far as where the satellite's binding energy equates that of the parent potential. Thus stars from lower-mass galaxies (which in turn have lower metallicity due to the effect of stellar winds) remain at larger radii, naturally creating a metallicity gradient.

Studies of metallicity profiles at very large radii are difficult, as the galaxy surface brightness becomes very low - orders of magnitude below the sky level. Nevertheless, innovative observations, such as deep long-slit integrations \cite[(Coccato et al., 2010)]{coccato10}, use of integral-field spectroscopy as a large single aperture \cite[(Weijmans et al., 2009, Greene et al., 2012)]{weijmans09,greene12}, or using multi-object spectroscopy to collect light over a 2D region \cite[(Foster et al., 2009)]{foster09} or from proxy tracers like globular clusters \cite[(Arnold et al., 2011)]{arnold11}, can measure line-strengths out to 3-4 effective radii (see Figure 3). In general, changes from the central gradient are subtle, and current small samples preclude general conclusions.

Colour profiles offer a promising technique to study these extremely faint outer regions. Deep imaging of massive galaxies often reveals direct signs of interactions \cite[(Duc et al., 2011)]{duc11}. Stacking carefully selected shallower data can also probe depths out to 8 effective radii \cite[(La Barbera et al., 2012)]{labarbera12} - much further out than is possible with current spectroscopic instruments. In addition to the difficulties of accurate sky subtraction and flat fielding at these very low surface brightnesses, the effects of chromatic PSF differences is very important, as the faint halo of the PSF spreads low-level light across large scales, and varies with each filter. Secondly, the stellar population models suffer degeneracy between age and metallicity in broad-band photometry, and generally lack sensitivity to subtle metallicity changes. However, the extensive baseline in radius, and rich spatial information make deep imaging an important complement to spectroscopy.

\vspace{-0.5cm}
\section{Environment}

Much has been written on the role of environment in influencing galaxy evolution, and, in particular, the star-formation process within galaxies. Rich clusters are the largest over-densities of baryonic matter in the Universe, and in a hierarchical framework, assembled mass more quickly than their surroundings. But does the act of being in a cluster impose intrinsic stellar population differences on member galaxies compared to the field?

The influential study of \cite{thomas05} found a difference in the mean ages of massive galaxies in high-density and low-density regions independent of galaxy velocity dispersion. In a more recent study, using similar methods for a much larger sample of galaxies but with lower quality data, \cite{thomas10} revised this picture to one where the more massive population has stellar ages that are independent of environment. Instead, the difference is attributed to the significance of the low mass `rejuvenated' population, which are more common in low-density environments.

\begin{figure}[t]
\begin{center}
 \includegraphics[width=5.5in]{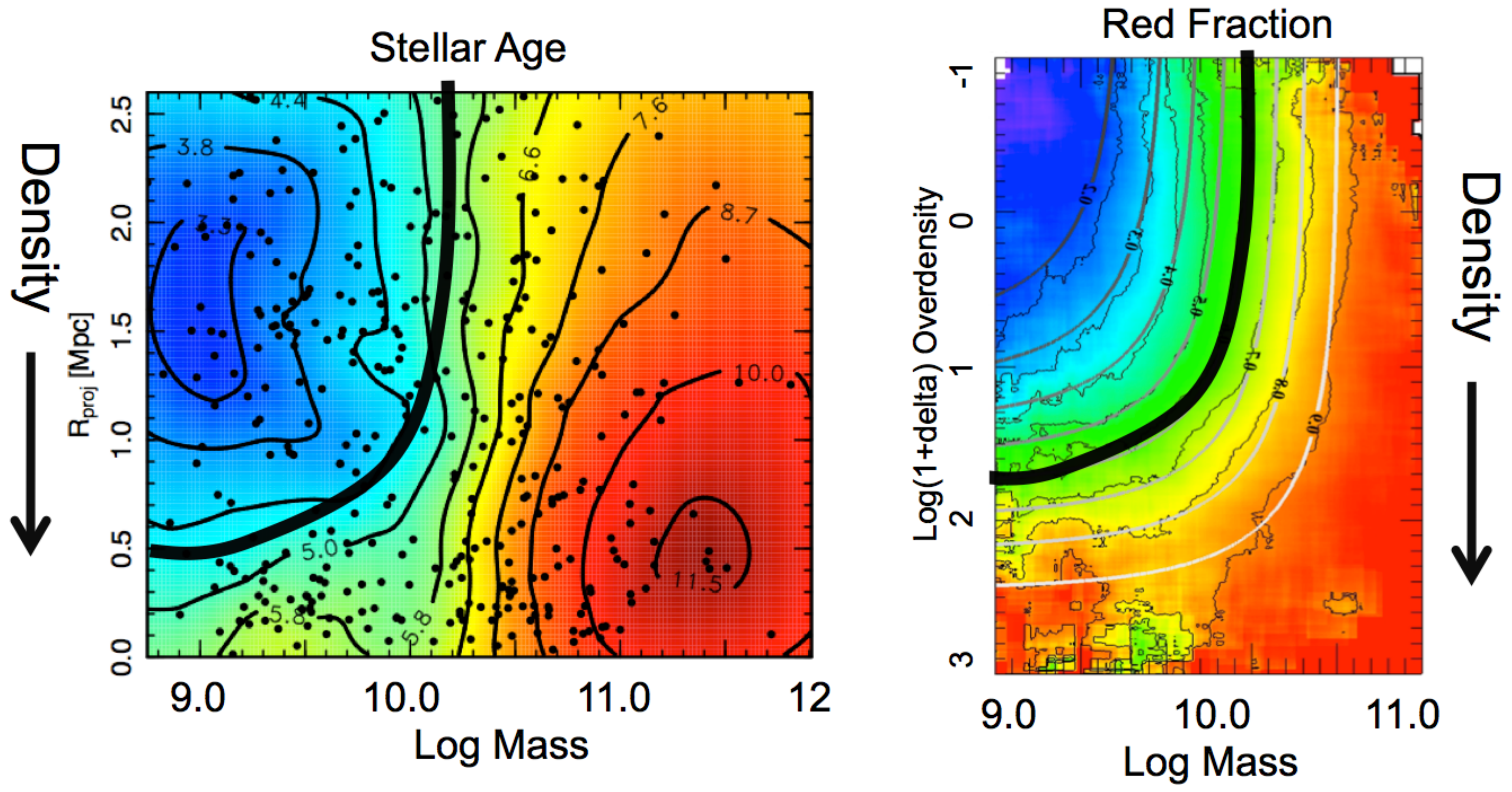} 
 \caption{{\it Left}: Stellar age versus mass and projected radial distance from the centre of the Coma cluster, from \cite{smith12a}. {\it Right}: Fraction of red galaxies versus mass and environmental density, from \cite{peng10}. The right-hand figure has been inverted to show density varying in the same direction in both figures. The matching thick hand-drawn line on both figures indicates the similar behavior of age and red fraction with mass and environmental density.}
   \label{fig4}
\end{center}
\end{figure}

A recent study by \cite{smith12a} addressed the question of environment using SDSS spectroscopy of the Coma cluster. 
This study found that, at high masses, the galaxy SSP-equivalent ages are only weakly sensitive to their position in the cluster, in the sense that all the most massive objects are old regardless of their position in the cluster. By contrast, the age of low-mass galaxies vary strongly with radius from the cluster centre. The combined effect is a reduction in the mean stellar age of galaxies with increasing distance from the cluster centre, which is mostly driven by the dependence of the lower-mass objects to the cluster-centric radius.

We show in Figure 4 that this picture is similar in essence to that of the apparent `separability' of mass and environment with respect to the quenching of star formation, proposed by \cite{peng10}. Whereas the specific mechanism for `mass quenching' is rather unclear, the fact that environment has an effect preferentially on lower mass systems is expected, given the clear signs of interactions and stripping that characterize the experience of a gas-rich galaxy falling into a larger parent or cluster halo \cite[(e.g. Vollmer et al., 2010)]{vollmer10}. 

As emphasized in \cite{smith12a}, the radial age dependence in Coma, driven by lower mass galaxies, can be understood when one considers that clusters are not well mixed and relaxed dynamical systems in the hierarchical paradigm. Indeed, for fixed halo mass, simulations show that the present-day position within the cluster is closely related (on average) to the infall epoch in the past \cite[(e.g. Smith et al., 2012a; De Lucia et al., 2012; McGee et al., 2012)]{smith12a,delucia12,mcgee09}. Thus environmental quenching associated to a galaxy transitioning from a `central' to a `satellite' object naturally imposes a temporal and spatial signature on the stellar population properties of cluster galaxies. Mass assembly is closely connected to this process via accretion and merging, and \cite{delucia12} therefore argue that the `history bias' inherent in the observational snapshot we see at any epoch may render the question of mass versus environment ill-posed.

\vspace{-0.5cm}
\section{Conclusions}

The main conclusions I wished to convey with this contribution are

\begin{itemize}
\item The `massive' galaxy population has a significant contribution from late-type spiral galaxies at all but the very highest masses. 
\item Compelling circumstantial evidence connects AGN activity with the transition of galaxies from the star-forming blue cloud to the passive red sequence. However, detailed studies of objects `caught in the act' of reducing their star-formation are needed to progress from correlation to causation.
\item Cold gas, young stars, and ongoing low-level star formation are frequently observed in early-type galaxies, and must be considered as part of their formation/evolution
\item The IMF varies systematically between low- and high-mass early-type galaxies. 
\item Spatially-resolved observations are a unique way to accurately connect observational features across widely different datasets. Spatial distributions and kinematic orientations also offer unique constraints on the origins of galaxy components.
\item The outer envelopes of galaxies are new frontiers for testing galaxy formation models
\item Environment plays a key role in our picture of galaxy formation and evolution. Insights from galaxy formation models help us to see these trends as part of a complete formation path, which may not easily be inferred from instantaneous view we  have.
\end{itemize}

\vspace{-0.5cm}
\section*{Acknowledgements}

It is a pleasure to thank the organizing committee, in particular the chairs Daniel Thomas, Anna Pasquali and Ignacio Ferreras, for the chance to contribute to a very enjoyable symposium. I would also like to thank my colleagues of the \atlas\/ consortium who's work I relied heavily on for this contribution. This work is supported by the Gemini Observatory, which is operated by the Association of Universities for Research in Astronomy, Inc., on behalf of the international Gemini partnership of Argentina, Australia, Brazil, Canada, Chile, the United Kingdom, and the United States of America.


\begin{thebibliography}{}
    \setlength{\baselineskip}{9.8pt}
\bibitem[Acquaviva et al.\,(2011)]{aquaviva11} Acquaviva, V., Gawiser, E., \& Guaita, L.\ 2011, ApJ, 737, 47 
\bibitem[Alatalo et al.\,(2011)]{alatalo11} Alatalo, K., Blitz, L., Young, L.~M., et al.\ 2011, ApJ, 735, 88 
\bibitem[Alatalo et al.\,(2012)]{alatalo12} Alatalo, K., Davis, T.~A., Bureau, M., et al.\ 2012, arXiv:1210.5524 
\bibitem[Arnold et al.\,(2011)]{arnold11} Arnold, J.~A., Romanowsky, A.~J., Brodie, J.~P., et al.\ 2011, ApJL, 736, L26 
\bibitem[Baldry et al.\,(2004)]{baldry04} Baldry, I.~K., Glazebrook, K., Brinkmann, J., et al.\ 2004, ApJ, 600, 681 
\bibitem[Baldry et al.\,(2006)]{baldry06} Baldry, I.~K., Balogh, M.~L., Bower, R.~G., et al.\ 2006, MNRAS, 373, 469 
\bibitem[Cappellari \& Emsellem\,(2004)]{cappellari04} Cappellari, M., \& Emsellem, E.\ 2004, PASP, 116, 138 
\bibitem[Cappellari et al.\,(2006)]{cappellari06} Cappellari, M., Bacon, R., Bureau, M., et al.\ 2006, MNRAS, 366, 1126 
\bibitem[Cappellari\,(2008)]{cappellari08} Cappellari, M.\ 2008, MNRAS, 390, 71 
\bibitem[Cappellari et al.\,(2011)]{cappellari11} Cappellari, M., Emsellem, E., Krajnovi{\'c}, D., et al.\ 2011, MNRAS, 413, 813
\bibitem[Cappellari et al.\,(2012a)]{cappellari12a} Cappellari, M., McDermid, R.~M., Alatalo, K., et al.\ 2012a, Nature, 484, 485 
\bibitem[Cappellari et al.\,(2012b)]{cappellari12b} Cappellari, M., McDermid, R.~M., Alatalo, K., et al.\ 2012b, arXiv:1208.3523 
\bibitem[Coccato et al.\,(2010)]{coccato10} Coccato, L., Gerhard, O., \& Arnaboldi, M.\ 2010, MNRAS, 407, L26 
\bibitem[Conroy \& van Dokkum\,(2012)]{conroy12} Conroy, C., \& van Dokkum, P.\ 2012, ApJ, 747, 69 
\bibitem[Davis et al.\,(2011)]{davis11} Davis, T.~A., Alatalo, K., Sarzi, M., et al.\ 2011, MNRAS, 417, 882 
\bibitem[Davis et al.\,(2012)]{davis12} Davis, T.~A., Krajnovi{\'c}, D., McDermid, R.~M., et al.\ 2012, MNRAS, 426, 1574
\bibitem[De Lucia et al.\,(2012)]{delucia12} De Lucia, G., Weinmann, S., Poggianti, B.~M., et al.\ 2012, MNRAS, 423, 1277 
\bibitem[Duc et al.\,(2011)]{duc11} Duc, P.-A., Cuillandre, J.-C., Serra, P., et al.\ 2011, MNRAS, 417, 863   
\bibitem[Emsellem et al.\,(1994)]{emsellem94} Emsellem, E., Monnet, G., \& Bacon, R.\ 1994, A\&A, 285, 723 
\bibitem[Faber et al.\,(2007)]{faber07} Faber, S.~M., Willmer, C.~N.~A., Wolf, C., et al.\ 2007, ApJ, 665, 265 
\bibitem[Ferreras et al.(2012)]{ferreras12} Ferreras, I., La Barbera, F., de Carvalho, R.~R., et al.\ 2012, arXiv:1206.1594 
\bibitem[Foster et al.\,(2009)]{foster09} Foster, C., Proctor, R.~N., Forbes, D.~A., et al.\ 2009, MNRAS, 400, 2135 
\bibitem[Gonz{\'a}lez\,(1993)]{gonzalez93} Gonz{\'a}lez, J.~J.\ 1993, Ph.D.~Thesis,  
\bibitem[Graves et al.(2009)]{graves09} Graves, G.~J., Faber, S.~M., \& Schiavon, R.~P.\ 2009, ApJ, 693, 486 
\bibitem[Greene et al.\,(2012)]{greene12} Greene, J.E., Murphy, J.D., Comerford, J.M., Gebhardt, K., \& Adams, J.J.\ 2012, ApJ, 750, 32 
\bibitem[Kauffmann et al.\,(2003)]{2003MNRAS.341...33K} Kauffmann, G., Heckman, T.~M., White, S.~D.~M., et al.\ 2003, MNRAS, 341, 33 
\bibitem[Kaviraj et al.\,(2007)]{kaviraj07} Kaviraj, S., Schawinski, K., Devriendt, J.~E.~G., et al.\ 2007, ApJS, 173, 619 
\bibitem[Kuntschner et al.\,(2010)]{kuntschner10} Kuntschner, H., Emsellem, E., Bacon, R., et al.\ 2010, MNRAS, 408, 97 
\bibitem[La Barbera et al.\,(2012)]{labarbera12} La Barbera, F., Ferreras, I., de Carvalho, R.~R., et al.\ 2012, MNRAS, 426, 2300 
\bibitem[McGee et al.\,(2009)]{mcgee09} McGee, S.~L., Balogh, M.~L., Bower, R.~G., et al.\ 2009, MNRAS, 400, 937 
\bibitem[Monnet et al.\,(1992)]{monet92} Monnet, G., Bacon, R., \& Emsellem, E.\ 1992, A\&A, 253, 366 
\bibitem[Morganti et al.\,(2006)]{morganti06} Morganti, R., de Zeeuw, P.~T., Oosterloo, T.~A., et al.\ 2006, MNRAS, 371, 157 
\bibitem[Ocvirk et al.\,(2006)]{ocvirk06} Ocvirk, P., Pichon, C., Lan{\c c}on, A., \& Thi{\'e}baut, E.\ 2006, MNRAS, 365, 74 
\bibitem[Oser et al.\,(2010)]{oser10} Oser, L., Ostriker, J.~P., Naab, T., Johansson, P.~H., \& Burkert, A.\ 2010, ApJ, 725, 2312 
\bibitem[Peng et al.\,(2010)]{peng10} Peng, Y.-j., Lilly, S.~J., Kova{\v c}, K., et al.\ 2010, ApJ, 721, 193 
\bibitem[Pforr et al.\,(2012)]{pforr12} Pforr, J., Maraston, C., \& Tonini, C.\ 2012, MNRAS, 422, 3285 
\bibitem[S{\'a}nchez-Bl{\'a}zquez et al.\,(2006)]{sanchez06} S{\'a}nchez-Bl{\'a}zquez, P., Peletier, R.~F., Jim{\'e}nez-Vicente, J., et al.\ 2006, MNRAS, 371, 703 
\bibitem[Schawinski et al.\,(2007)]{schawinski07} Schawinski, K., Thomas, D., Sarzi, M., et al.\ 2007, MNRAS, 382, 1415 
\bibitem[Scott et al.\,(2009)]{scott09} Scott, N., Cappellari, M., Davies, R.~L., et al.\ 2009, MNRAS, 398, 1835 
\bibitem[Serra et al.\,(2012)]{serra12} Serra, P., Oosterloo, T., Morganti, R., et al.\ 2012, MNRAS, 422, 1835 
\bibitem[Shapiro et al.\,(2010)]{shapiro10} Shapiro, K.~L., Falc{\'o}n-Barroso, J., van de Ven, G., et al.\ 2010, MNRAS, 402, 2140 
\bibitem[Smith et al.\,(2012a)]{smith12a} Smith, R.~J., Lucey, J.~R., Price, J., Hudson, M.~J., \& Phillipps, S.\ 2012a, MNRAS, 419, 3167 
\bibitem[Smith et al.\,(2012b)]{smith12b} Smith, R.~J., Lucey, J.~R., \& Carter, D.\ 2012b, MNRAS, 426, 2994 
\bibitem[Spiniello et al.(2012)]{spiniello12} Spiniello, C., Trager, S.~C., Koopmans, L.~V.~E., \& Chen, Y.~P.\ 2012, ApJL, 753, L32
\bibitem[Thomas et al.\,(2005)]{thomas05} Thomas, D., Maraston, C., Bender, R., \& Mendes de Oliveira, C.\ 2005, ApJ, 621, 673 
\bibitem[Thomas et al.\,(2010)]{thomas10} Thomas, D., Maraston, C., Schawinski, K., Sarzi, M., \& Silk, J.\ 2010, MNRAS, 404, 1775 
\bibitem[Trager et al.\,(2000)]{trager00} Trager, S.~C., Faber, S.~M., Worthey, G., \& Gonz{\'a}lez, J.~J.\ 2000, AJ, 119, 1645 
\bibitem[Treu et al.\,(2010)]{treu10} Treu, T., Auger, M.~W., Koopmans, L.~V.~E., et al.\ 2010, ApJ, 709, 1195 
\bibitem[van Dokkum \& Conroy\,(2010)]{vandokkum10} van Dokkum, P.~G., \& Conroy, C.\ 2010, Nature, 468, 940 
\bibitem[Vazdekis et al.\,(2012)]{vazdekis12} Vazdekis, A., Ricciardelli, E., Cenarro, A.~J., et al.\ 2012, MNRAS, 424, 157 
\bibitem[Vollmer et al.\,(2010)]{vollmer10} Vollmer, B., Soida, M., Chung, A., et al.\ 2010, A\&A, 512, A36 
\bibitem[Weijmans et al.\,(2009)]{weijmans09} Weijmans, A.-M., Cappellari, M., Bacon, R., et al.\ 2009, MNRAS, 398, 561 
\bibitem[Young et al.(2008)]{young08} Young, L.~M., Bureau, M., \& Cappellari, M.\ 2008, ApJ, 676, 317 
\bibitem[Young et al.\,(2011)]{young11} Young, L.~M., Bureau, M., Davis, T.~A., et al.\ 2011, MNRAS, 414, 940 

\end{thebibliography}
\end{document}